\documentclass[10pt,twocolumn,twoside]{osajnl}

\journal{ol} 

\setboolean{shortarticle}{true} 

\title{Screening and fluctuation of topological charge in random wave fields}

\author[1]{L. De Angelis}
\author[1,*]{L. Kuipers}

\affil[1]{Kavli Institute of Nanoscience, Delft University of Technology, 2600 GA, Delft, The Netherlands}

\affil[*]{Corresponding author: l.kuipers@tudelft.nl}

\ociscodes{(260.6042) Singular optics; (180.4243) Near-field microscopy; (140.1540) Chaos.}

\begin{abstract}
Vortices, phase singularities and topological defects of any kind
often
reflect information that is crucial to understand
physical systems in which such entities arise.
With near-field experiments supported by numerical calculations, we
determine the fluctuations of the topological charge for phase singularities
in isotropic random waves, as a function of the size $R$ of the
observation window. We demonstrate that for 2D fields such fluctuations increase
with a super-linear scaling law, consistent with a $R\log{R}$ behavior. Additionally, we show that
such scaling remains valid
in presence of anisotropy.
\end{abstract}

\setboolean{displaycopyright}{true}

\begin{document}

\maketitle

An accurate knowledge on the statistical fluctuations of a given
physical observable is often essential, with an importance
on par with the ensemble-averaged value of the observable itself~\cite{Blibar2008}.
In fact, fluctuations are ubiquitous in quantum as well as in
classical physics.
A prime
example is the grand canonical ensemble in statistical physics,
where the number of particles is only known in average, and its fluctuations
have an actual physical meaning, directly linked to the chemical potential of the system~\cite{Frenkel2001}.
Ensembles of this type are offered by many systems in physics,
for example whenever they exhibit topological defects~\cite{Genkin2017,Prokhorenko2017},
which in the context
of optics can be optical singularities~\cite{Nye1974}. These singularities
are point-like entities carrying a topological charge and in random waves they
are reminiscent of interacting particles~\cite{DeAngelis2016}. Oppositely
charged pairs can be created and destroyed~\cite{Cheng2014}, resulting in a total number of singularities
which is not conserved. Although the total topological charge of an ensemble of singularities
is always conserved~\cite{Gbur2016}, this number can still vary when considering a
finite observation window, and its fluctuations
are the hallmark for intrinsic properties of the system, such as charge screening~\cite{VanTiggelen2006}.

Here we study quantitatively the fluctuations of the total topological charge for
phase singularities in random waves and determine the dependence of such
fluctuations on the size of the observation window.
With near-field experiments we map the optical near-field inside a chaotic
cavity~\cite{DeAngelis2016}. By tuning the excitation
wavelength we measure different
realizations of the optical random wave pattern inside the cavity~\cite{DeAngelis2017}.
Such phase- and polarization-resolved measurements enable us to pinpoint position
and topological charge of the individual phase singularities in all in-plane
components of the electric field that we measure, and therefore determine and investigate
their total
topological charge and its fluctuations.
With experimental evidence, corroborated by numerical calculations,
we demonstrate that the sum of the topological charges contained in a square
region of area $R^2$ fluctuates as $R\log{R}$, in agreement with
analytical calculations~\cite{VanTiggelen2006}.

\begin{figure*}[t!]
	\centering
	\includegraphics[width = \textwidth]{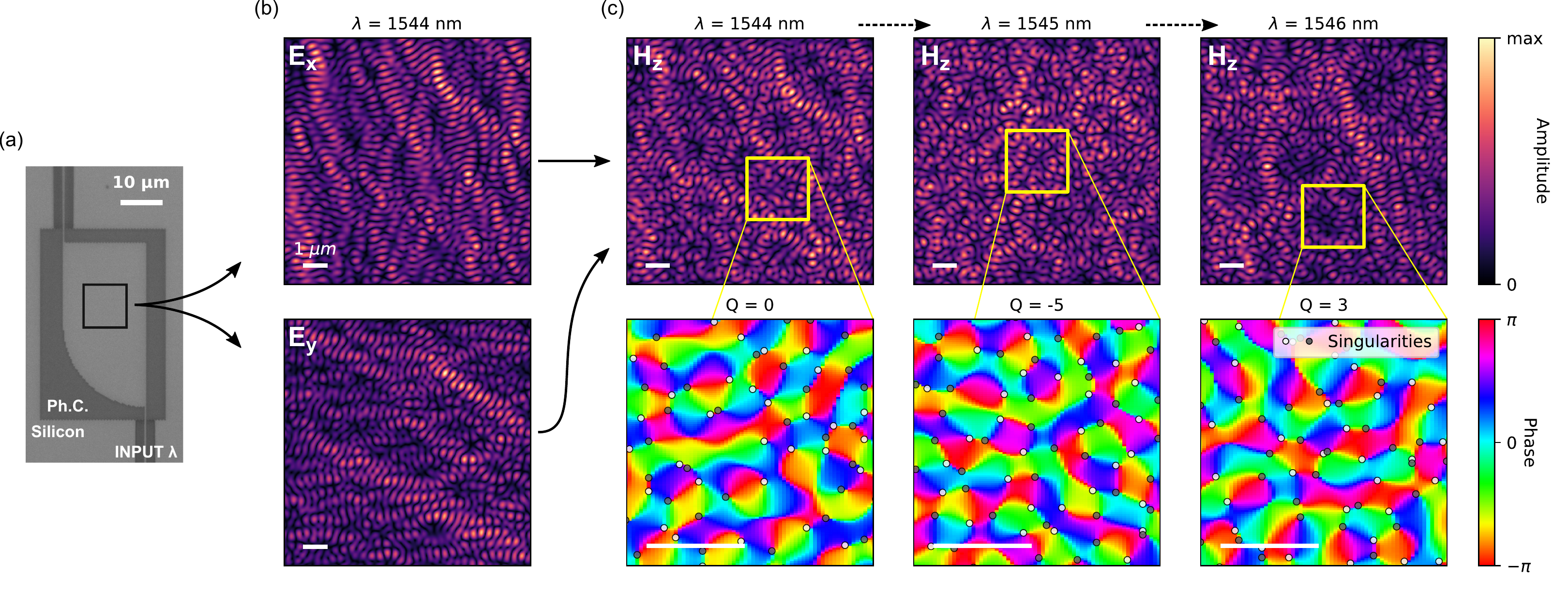}
	\caption{Overview of the near-field measurements of
		the optical field inside the chaotic cavity.
		(a) Optical micrograph of the cavity used for the generation of the optical
		random wave field. The dark area is the
		photonic crystal that confines light inside the cavity. 
		(b) Example of a direct measurement of the in-plane components of the optical
		random field under investigation. In the upper panel the amplitude of $E_x$ and
		in the lower pane the amplitude of $E_y$.
		(c) Near-field maps of $H_z\propto\mathbf{k}\times\mathbf{E}$.
		In the upper panels the amplitude of $H_z$ for different excitation wavelength
		$\lambda$, and in the lower panels zoomed-in images of its phase.
		Phase singularities of positive and negative topological charge are depicted
		by light-gray and dark-gray circles, respectively.}
	\label{fig:overview}
\end{figure*}

We generate optical random waves by coupling infrared
monochromatic light into a chaotic 
cavity [Fig.~\ref{fig:overview}(a)]. This consists
of a 220~nm silicon membrane on a silica buffer, patterned
with a photonic crystal which encloses the cavity area.
The shape of this area was engineered so
to ensure random wave propagation
in the cavity~\cite{Liu2015,Stockmann2006}.
By means of near-field optical microscopy we map amplitude,
phase and polarization of the in-plane
optical field inside the cavity~\cite{Rotenberg2014}.
While our measurements provide access to
the two-dimensional random vector field,
previous theoretical results describe the behavior of
a scalar quantity~\cite{VanTiggelen2006}.
With this regard, it is important to note that
our complete information on the in-plane field $\mathbf{E}$
allows us to reconstruct an out-of-plane component
$H_z\propto \mathbf{k}\times\mathbf{E}$, which behaves fully as a
scalar field~\cite{DeAngelis2016}.
Following well established models for random wave fields~\cite{Stockmann2006},
we can think of $H_z$ as an
isotropic superposition of plane waves interfering with random phases $\phi_\mathbf{k}$~\cite{Berry2000},
\begin{equation}
H_z = \sum_{|\mathbf{k}| = k_0}\exp({i\mathbf{k}\cdot\mathbf{r} +i \phi_\mathbf{k}}),
\label{eq:rwaves}
\end{equation}
which is characterized by an autocorrelation
\begin{equation}
C(r) = \int d\mathbf{r}' H_z(\mathbf{r}') H_z(\mathbf{r}+\mathbf{r}') = J_0(kr),
\label{eq:rwaves}
\end{equation}
where $J_0(kr)$ is the Bessel function of order 0.

Figure \ref{fig:overview} presents a direct measurement of the amplitude of the
in-plane field components ($E_x,E_y$), as well as amplitude and phase of
$H_z$, as obtained from measurements at different excitation wavelength.
From the subwavelength maps of the amplitude
we can clearly resolve the interference that results
in a speckle-like pattern~\cite{Yilmaz2015}. Figure~\ref{fig:overview} also displays
zoomed-in images for the phase of $H_z$. Here, the circles indicate the location
of phase singularities with their topological charge (color), i.e.,
the integer number of times that the phase of the field
loops from $-\pi$ to $\pi$ around the singular point.
We pinpoint the location of each singularity by integrationg the phase
variations at the experimental limit of two by two pixels, which
sets our spatial resolution to be at best equal to the pixel size,
of approximately $20$ nm.
We always observe the topological charge to be
$\pm 1$ (dark/light gray)~\cite{Nye1974}.
The patterns presented in Fig.~\ref{fig:overview} change dramatically with
the input wavelength. A wavelength shift of 1 nm already leads to a totally
different field configuration. In fact, the spectral correlation width of this
random field is of the order of 0.2 nm, as we
quantify by computing the wavelength-wavelength correlation of $H_z$ \cite{DeAngelis2017}.

Although the wave field is made up by randomly interfering waves,
the distribution of the singularities does contain structure.
In fact, the distribution of phase singularities in random waves
has a liquid-like correlation~\cite{Berry2000,Stockmann2009,DeAngelis2016}.
An immediate question that arises at this point, is whether
the charges of such distribution of singularities are correlated
or not, and, if so, how. In a system of charged particles we would expect
such correlation to occur due to charge screening.
While it is tempting to make an analogy straight away,
and predict a screening among topological charges,
we must remember that the nature of phase
singularities is radically different from that of atoms and molecules,
and there is no true and measurable physical interaction among these entities.
A straightforward analogy between charged particles and singularities with their
topological charge is therefore not so trivial.

\begin{figure*}[t]
	\centering
	\includegraphics[width = 16cm]{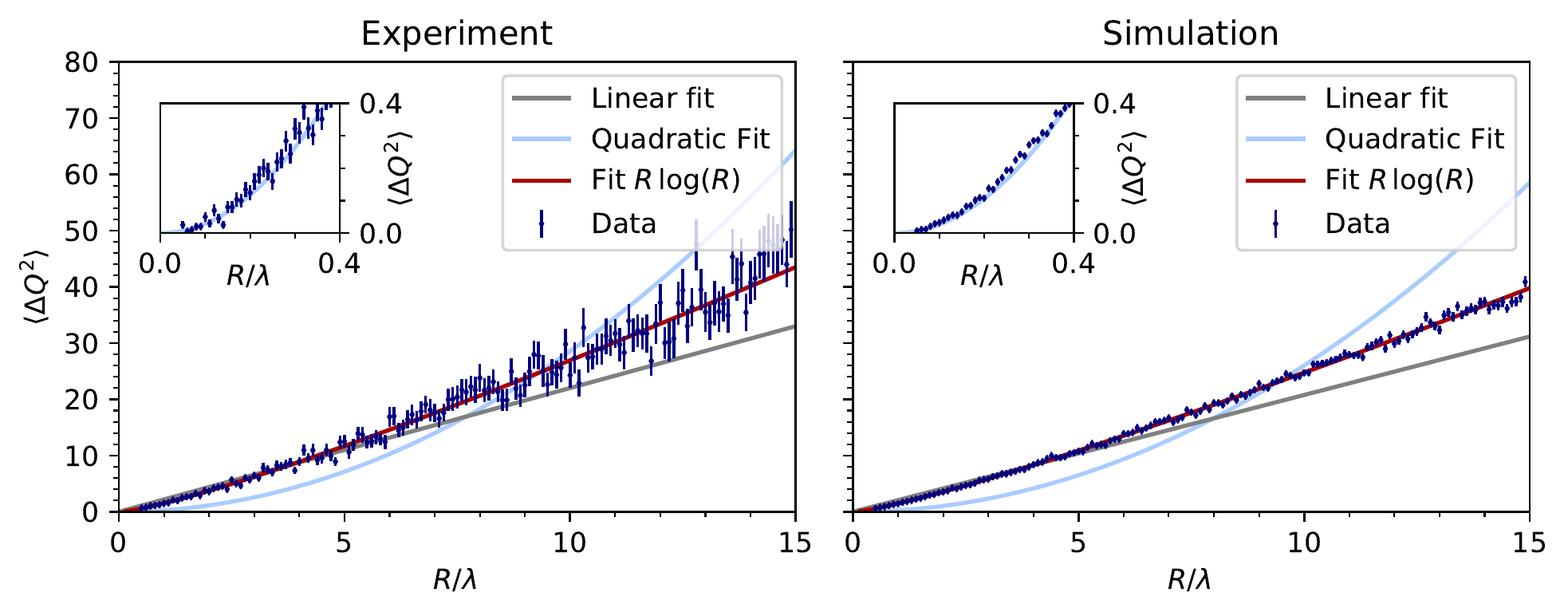}
	\caption{Fluctuation of topological charge $\langle\Delta Q^2\rangle$
		as a function of the size $R$ of the observation area $R^2$,
		for the experimental scalar field $H_z$ (a) and for a
		simulated isotropic scalar field (b).
		The blue errorbars
		are the respective data points.
		The gray lines are the best fit with
		$f_1 = mR$ [$m_{exp} = 2.20(5)$; $m_{sim}=2.08(4)$]. The red lines are the best fit with
		$f_2 = a\,bR\log{bR}$ [$a_{exp} = 0.026(5)$, $b_{exp} = 19(3)$; $a_{sim}=0.0144(1)$, $b_{sim} = 30(2)$].
		The light blue lines are the best fit with
		$f_3 = cR^2$ [$c_{exp} = 0.286(11)$; $c_{sim}=0.260(11)$].}
	\label{fig:scalar_results}
\end{figure*}

The easiest test that can be performed to determine the existence of
charge screening,
is to measure the overall topological charge $Q=\sum_{i}q_i$ of the
singularities $q_i$ contained in an area
of dimension $A=R^2$. In complete absence of charge correlation
one expects the average $\langle Q\rangle$ of such quantity to
be zero, and its variance $\langle Q^2\rangle$ to scale with the area
of the observation window $R^2$~\cite{VanTiggelen2006}.
A screening among charges would reveal itself
by slowing down the dependence of such variance to a sub-quadratic law.
In fact, screening neutralizes charges by surrounding them with a cloud of opposite charges,
so to prevent fluctuations of the total charge inside an area $\sim R^2$,
in favor of fluctuations along the perimeter region $\sim R$~\cite{Egorov2008}.
The existence of screening among topological charges
is well established in literature. \cite{Freund1998,Dennis2003,VanTiggelen2006,Egorov2008,Kessler2008JOSA,Kessler2008OC,Freund2015,Freund2007,Dennis2017}.
It starts to play a role when the size of the
observation window is bigger than the typical inter-singularity distance,
of approximately $\lambda/2$.
However, how much this screening slows
down the quadratic law $\langle Q^2(R)\rangle \propto R^2$
is yet unclear.
Explicitly, a first model of $\langle Q^2(R)\rangle$, in which two assumptions were made on the
autocorrelation of the random field, predicted linear scaling~\cite{Freund1998},
whereas
further theory developments proved such quantity to scale as $R\log{R}$~\cite{VanTiggelen2006}.
More recently, also paraxial experiments were performed, supporting the linear
dependence~\cite{Freund2007,Egorov2008,Freund2015}.

Figure~\ref{fig:scalar_results} presents our results for
$\langle Q^2(R)\rangle$. In the main plots the
analysis of screening for $R\geq \lambda/2$,
in the inset a proof of its absence in the region $R<\lambda/2$.
In the left panel
we present the experimental data, which is the result of the sampling of 200
experimental scalar field realizations
where we randomly pick the position of our observation window.
The fields are obtained by varying
the excitation wavelength $\lambda$ over a range $\Delta\lambda = 20$~nm around $\lambda_0 = 1550$~nm.
In the right panel we show simulation data, realized by sampling 3500 random wave fields
which were
independently calculated by adding up
250 plane waves with isotropic directions and
random phases [\eqref{eq:rwaves}]. In both cases, we obtain a good agreement only with
the $R\log{R}$ dependence. A quadratic fit $c R^2$
is clearly inadequate for the data displayed in the main plot.
However, this functional behavior perfectly describes
the short-distance data in the insets of Fig.~\ref{fig:scalar_results}, at which range screening is indeed absent.
Less evident, but still significant, is the inconsistency between
the data and a linear behavior. Although this
is more eye-catching in the region $R/\lambda > 5$, a clear
deviation is still present for $R/\lambda < 3$,
with even greater significance considering the small errorbars
associated with the latter region.

More quantitatively, we fitted our data with
$f_1(R) = mR$ (gray lines in Fig.~\ref{fig:scalar_results}),
and $f_2(R) = a\,bR\log{bR}$ (red lines in Fig.~\ref{fig:scalar_results}).
We focus on how well these functions can describe our data
rather than on the resulting fit parameters, which
we found to depend on the shape of the chosen observation window (not shown).
We quantified the goodness of such least-square fits by performing a
$\chi^2$ test, the results of which are summarized in Table~\ref{tab:chi2_scal}.
In both experiment and simulation we are performing the fit
on 145 equally spaced data points, resulting in 144 degrees of freedom (DOF)
for the linear fit (1 free parameter) and in 143 DOF for $f_2$ (2
free parameters). The $\chi^2$ is a stochastic variable, with expectation
value equal to the DOF~\cite{Cochran1952}.
The values of the $\chi^2$ for the fittings with $f_2(R)$
are consistent with their expectation value, whereas
the case of $f_1$
leads to $\chi^2$ values that are too high to be mere statistical fluctuations.
After this quantitative analysis of our fits we can most certainly
conclude that the $R\log R$ scaling law is describing the behavior
of $\langle Q^2(R) \rangle$ better than a linear function.

\begin{table}[h]
	\centering
	\begin{tabular}{cccc}
		\hline\hline
		Fit function	& ${\chi^2}_{exp}$	&	 ${\chi^2}_{sim}$	&	DOF		\\ \hline
		$f_1(R) = mR$	&	954				&	     $10^4$			&	144		\\
		$f_2(R) = a\,bR\log{bR}$ &	125			&			147			&	143 \\ \hline\hline
	\end{tabular}
	\caption{Summary of the $\chi^2$ tests for the least-square fits
		presented in Fig.~\ref{fig:scalar_results} (isotropic case).}
	\label{tab:chi2_scal}
\end{table}

Certainly, the studied cases are not exhaustive of all the possible
functional behaviors one could think of. For instance, an alternative
trade-off between the linear and quadratic scalings 
could be given by a generic power law $\beta R^\alpha$.
Interestingly, such a function can be effectively
used to fit both experimental and simulated data,
with $\alpha\approx 1.2$.
However, the result of such fits (not shown) are found to be less
reliable, since they lead to different optimal fit parameters when
varying fitting range.
In the absence of existing theories,
they remain difficult to interpret.

Going back to screening and its nature, we now investigate its role
in presence of anisotropy. In fact, in case of
anisotropic wave propagation also the
spatial arrangement of phase singularities becomes anisotropic~\cite{DeAngelis2016},
and it does not resemble the distribution of a simple liquid anymore.
Actually, along particular directions the resulting distribution is
more reminiscent of an ordered structure.
Thus, it is interesting to check whether or not
this anisotropy influences the topological screening here discussed.

Anisotropic wave propagation naturally takes place
in the single vector components of the measured in-plane electric field.
This is caused by the strict relation between polarization and propagation direction
set by transverse electric propagation~\cite{DeAngelis2016}.
Figure~\ref{fig:overview}(b) presents an example of our direct
measurement of amplitude $E_x$ and $E_y$ inside the chaotic cavity.
By comparing this figures to the maps for the scalar field $H_z$
presented in Fig.~\ref{fig:overview}(c), we can see a pronounced anisotropy.
For example, in the amplitude map of $E_x$ we easily distinguish a stripy pattern, given by a fast
modulation
of the amplitude along the $y$-axis, opposed to a modulation along the
$x$-axis which is slower by at least a factor two.
This anisotropy results in a spatial arrangement
of dislocations where many
singularities with the same topological charge are displaced along the $y$-axis,
while the first neighbor in the $x$-direction is often oppositely charged~\cite{DeAngelis2016}.

\begin{figure}[h]
	\centering
	\includegraphics[width = \linewidth]{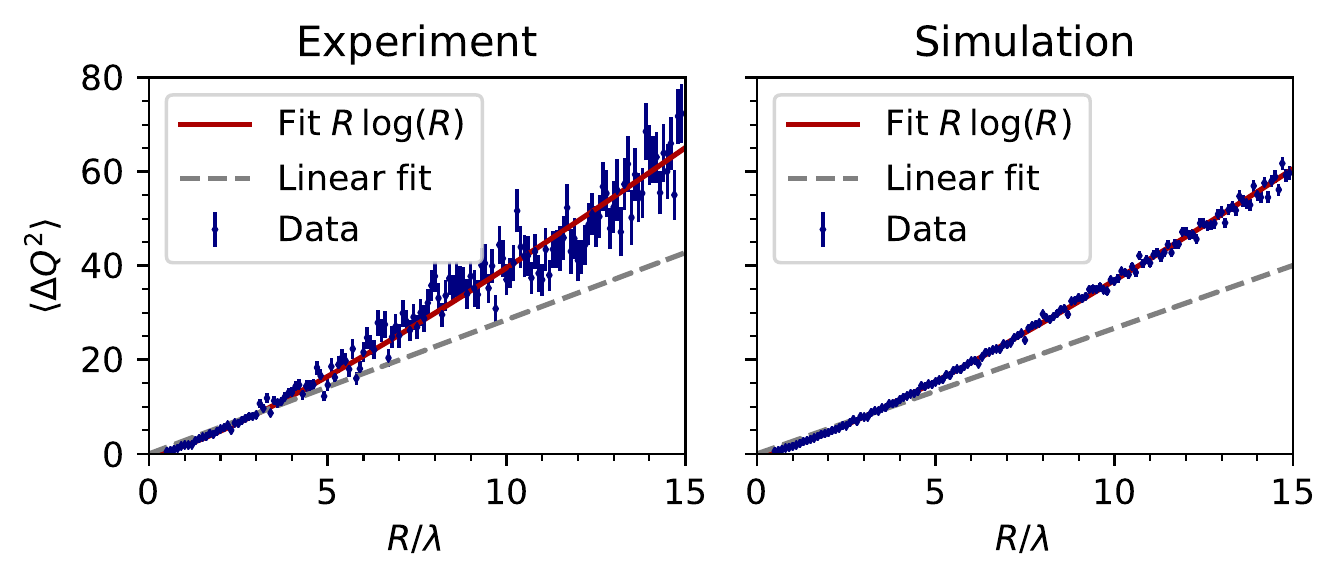}
	\caption{Fluctuation of the topological charge $\langle\Delta Q^2\rangle$
		vs. $R$, as in Fig.~\ref{fig:scalar_results}.
		These results are for the anisotropic case of a Cartesian
		component ($E_x$) of a trasverse vector field $\mathbf{E}$.
		Gray lines
		$f_1 = mR$ [$m_{exp} = 2.85(9)$; $m_{sim}=2.08(4)$]. Red lines 
		$f_2 = a\,bR\log{bR}$ [$a_{exp} = 0.15(2)$, $b_{exp} = 6.5(5)$; $a_{sim}=0.126(3)$, $b_{sim} = 6.9(1)$].}
	\label{fig:vec_results}
\end{figure}

Figure~\ref{fig:vec_results} presents the fluctuation of topological
charge $\langle\Delta Q^2\rangle$ for the case of $E_x$, in both experiment
and simulation. The analysis of its behavior is carried in complete analogy to what
already described for $H_z$. Again, we can conclude that the scaling
law given by $R\log{R}$ is more successful than a linear function
[$\chi^2_{exp}(mR) = 2098$ vs  $\chi^2_{exp}(a\,bR\log{bR}) = 187$].
However, we do observe that the growth rate of $\langle\Delta Q^2\rangle$
is faster than in the case of
the scalar field $H_z$ (Fig.~\ref{fig:scalar_results}).
This suggests that an anisotropic distribution of topological charges
results in a screening that on average is less effective with respect to its
isotropic counterpart. This of course only holds when considering the
average over all the possible directions along which singularities are displaced, whereas it is
very likely for this form of screening to strongly depend on the considered direction.
Still, considering these qualitative differences with
the isotropic scalar case,
it is remarkable how the $R\log{R}$ law can still describe the data.

To conclude, 
we presented a quantitative study on
the screening of topological charges for singularities in random waves.
With near-field experiments and numerical calculations we demonstrated that while
the average topological charge remains zero independently of the size $R$ of the observation window,
the fluctuation of this quantity increases with a dependence which is consistent with a $R\log{R}$ law.
This result validates previous analytical
theory~\cite{VanTiggelen2006}. Additionally, we extended our study to the
case of anisotropic wave propagation, and even though the nature of
screening drastically changes, we showed that
the functional dependence of topological charge fluctuations is still
well described by the  $R\log{R}$ scaling law.

We thank Andrea Di Falco for fabricating the chaotic cavity used in the near-field experiments,
Su-Hyun Gong for critical reading of the manuscript and Filippo Alpeggiani for useful discussions.
This work is part of the research
program of the Netherlands Organization for Scientific Research (NWO).
The authors acknowledge
funding from the European Research Council (ERC Advanced Grant
No. 340438-CONSTANS).

\bibliography{screening_ref}

\begin{thebibliography}{10}
\newcommand{\enquote}[1]{``#1''}

\bibitem{Blibar2008}
S.~Balibar, Nature \textbf{451}, 136 (2008).

\bibitem{Frenkel2001}
D.~Frenkel and B.~Smit, \emph{Understanding molecular simulation: from
  algorithms to applications}, vol.~1 (Academic press, 2001).

\bibitem{Genkin2017}
M.~M. Genkin, A.~Sokolov, O.~D. Lavrentovich, and I.~S. Aranson, Phys. Rev. X
  \textbf{7}, 011029 (2017).

\bibitem{Prokhorenko2017}
S.~Prokhorenko, Y.~Nahas, and L.~Bellaiche, Phys. Rev. Lett. \textbf{118},
  147601 (2017).

\bibitem{Nye1974}
J.~F. Nye and M.~V. Berry, Proc. R. Soc. Lond. A \textbf{336}, 165 (1974).

\bibitem{DeAngelis2016}
L.~De~Angelis, F.~Alpeggiani, A.~Di~Falco, and L.~Kuipers, Phys. Rev. Lett.
  \textbf{117}, 093901 (2016).

\bibitem{Cheng2014}
X.~Cheng, Y.~Lockerman, and A.~Z. Genack, Opt. Lett. \textbf{39}, 3348 (2014).

\bibitem{Gbur2016}
G.~Gbur, Optica \textbf{3}, 222 (2016).

\bibitem{VanTiggelen2006}
B.~A. van Tiggelen, D.~Anache, and A.~Ghysels, Europhys. Lett. \textbf{74}, 999
  (2006).

\bibitem{DeAngelis2017}
L.~De~Angelis, F.~Alpeggiani, A.~Di~Falco, and L.~Kuipers, Phys. Rev. Lett.
  \textbf{119}, 203903 (2017).

\bibitem{Liu2015}
C.~Liu, R.~E.~C. van~der Wel, N.~Rotenberg, L.~Kuipers, T.~F. Krauss, A.~{Di
  Falco}, and A.~Fratalocchi, Nat. Phys. \textbf{11}, 358 (2015).

\bibitem{Stockmann2006}
H.-J. St{\"o}ckmann, \emph{Quantum chaos: an introduction} (Cambridge
  university press, Cambridge, 2006).

\bibitem{Rotenberg2014}
N.~Rotenberg and L.~Kuipers, Nat. Photonics \textbf{8}, 919 (2014).

\bibitem{Berry2000}
M.~V. Berry and M.~R. Dennis, Proc. R. Soc. Lond. A \textbf{456}, 2059 (2000).

\bibitem{Yilmaz2015}
H.~Yilmaz, E.~G. van Putten, J.~Bertolotti, A.~Lagendijk, W.~L. Vos, and A.~P.
  Mosk, Optica \textbf{2}, 424 (2015).

\bibitem{Stockmann2009}
R.~H\"ohmann, U.~Kuhl, H.-J. St\"ockmann, J.~D. Urbina, and M.~R. Dennis, Phys.
  Rev. E \textbf{79}, 016203 (2009).

\bibitem{Egorov2008}
R.~I. Egorov, M.~S. Soskin, D.~A. Kessler, and I.~Freund, Phys. Rev. Lett.
  \textbf{100}, 103901 (2008).

\bibitem{Freund1998}
I.~Freund and M.~Wilkinson, J. Opt. Soc. Am. A \textbf{15}, 2892 (1998).

\bibitem{Dennis2003}
M.~R. Dennis, J. Phys. A-Math. Gen. \textbf{36}, 6611 (2003).

\bibitem{Kessler2008JOSA}
D.~A. Kessler and I.~Freund, J. Opt. Soc. Am. A \textbf{25}, 2932 (2008).

\bibitem{Kessler2008OC}
I.~Freund and D.~A. Kessler, Opt. Commun. \textbf{281}, 5954  (2008).

\bibitem{Freund2015}
I.~Freund, D.~A. Kessler, V.~Vasyl'ev, and M.~S. Soskin, Opt. Lett.
  \textbf{40}, 4747 (2015).

\bibitem{Freund2007}
I.~Freund, R.~I. Egorov, and M.~S. Soskin, Opt. Lett. \textbf{32}, 2182 (2007).

\bibitem{Dennis2017}
A.~J.~H. Houston, M.~Gradhand, and M.~R. Dennis, J. Phys. A-Math. Theor.
  \textbf{50}, 205101 (2017).

\bibitem{Cochran1952}
W.~G. Cochran, Ann. Math. Stat. \textbf{23}, 315 (1952).

\end{thebibliography}

\bibliographyfullrefs{screening_ref}

\end{document}